\documentclass[conference]{IEEEtran}
\IEEEoverridecommandlockouts
\usepackage{cite}
\usepackage{amsmath,amssymb,amsfonts}
\usepackage{algorithmic}
\usepackage{graphicx}
\usepackage{textcomp}
\usepackage{xcolor}
\usepackage{soul}
\def\BibTeX{{\rm B\kern-.05em{\sc i\kern-.025em b}\kern-.08em
    T\kern-.1667em\lower.7ex\hbox{E}\kern-.125emX}}

\usepackage{url}
\usepackage{siunitx}
\usepackage{subcaption}
\usepackage{balance}
\usepackage{gensymb}
\usepackage[acronym]{glossaries}
\makeglossaries
\newacronym{ofdm}{OFDM}{Orthogonal Frequency Division Multiplexing}
\newacronym{mimo}{MIMO}{Multiple-Input Multiple-Output}
\newacronym{crlb}{CRLB}{Cramér-Rao Lower Bound}
\newacronym[plural=SAFs]{saf}{SAF}{Spatial Ambiguity Function}
\newacronym{csi}{CSI}{Channel State Information}
\newacronym{los}{LoS}{Line-of-Sight}
\newacronym{ue}{UE}{User Equipment}
\newacronym{cfo}{CFO}{Carrier Frequency Offset}
\newacronym{ula}{ULA}{Uniform Linear Array}
\newacronym{pmsr}{PMSR}{Peak-to-Median Sidelobe Ratio}
\newacronym{fim}{FIM}{Fisher Information Matrix}
\newacronym{efim}{EFIM}{Effective FIM}
\newacronym{rmse}{RMSE}{Root Mean Squared Error}
\newacronym{lse}{LSE}{Least Squares Estimator}
\newacronym{isac}{ISAC}{Integrated Sensing and Communication}
\newacronym{mocap}{MoCap}{Motion Capture}
\newacronym{fft}{FFT}{Fast Fourier Transform}
\newacronym[
  plural=Txs,
  firstplural=transmitters (Txs)
]{tx}{Tx}{transmitter}
\newacronym[
  plural=Rxs,
  firstplural=receivers (Rxs)
]{rx}{Rx}{receiver}
\newacronym{cpi}{CPI}{Coherent Processing Interval}
\newacronym{mle}{MLE}{Maximum Likelihood Estimator}
\newacronym{4d}{4D}{four-dimensional}
\newacronym{ff}{FF}{far-field}
\newacronym{nf}{NF}{near-field}
\newacronym{2d}{2D}{two-dimensional}
\newacronym{snr}{SNR}{Signal to Noise Ratio}

\usepackage[font=footnotesize,labelfont=bf]{caption}
\begin{document}

\title{
Near-Field Velocity Estimation and Doppler-Aware Localization in OFDM Massive MIMO
\thanks{
This work is supported by the EU HORIZON-MSCA-DN-2022 No. 101119652 (6\textsuperscript{th}Sense), Horizon Europe Research and Innovation programme No. 101192521 (MultiX), and No. 101139257 (SUNRISE-6G).
}
}

\author{
  Qing Zhang\textsuperscript{*,\ensuremath{\dagger}},
  Dario Tagliaferri\textsuperscript{\ensuremath{\ddagger}},
  Robbert Beerten\textsuperscript{*},
  Zhuangzhuang Cui\textsuperscript{*,\ensuremath{\dagger}},
  Yang Miao\textsuperscript{\#}, 
  and Sofie Pollin\textsuperscript{*,\ensuremath{\dagger}} \\
  \textsuperscript{*}Department of Electrical Engineering (ESAT), KU Leuven, Belgium \\
  \textsuperscript{\ensuremath{\dagger}}Interuniversity Microelectronics Centre (IMEC), Belgium\\
  \textsuperscript{\ensuremath{\ddagger}}Department of Electronics, Information, and Bioengineering (DEIB), Politecnico di Milano, Italy \\
  \textsuperscript{\#}Department of Electrical Engineering (EEMCS-EE), University of Twente, The Netherlands
  \vspace{-0.4cm}
}

\maketitle

\begin{abstract}
In \gls{ofdm}-based massive \gls{mimo} \gls{nf} sensing, target motion induces an antenna-dependent bistatic Doppler variation across the array aperture. Ignoring this spatial Doppler variation leads to a model mismatch that degrades \gls{nf} localization. In this paper, we propose a low-complexity recursive framework for joint radial/transverse velocity estimation and Doppler-aware localization. Initialized by a constant-Doppler coarse localization, the method alternates between closed-form \gls{lse}-based velocity estimation and antenna-dependent Doppler-aware localization refinement. Simulation and measurement results demonstrate the effectiveness of the proposed framework against two benchmark methods. Compared with a low-complexity constant-Doppler baseline method, the proposed algorithm improves range, angle, and radial velocity estimation results, while also enabling transverse velocity estimation. In the measurement results, the overall localization error decreases from \(0.268\;\mathrm{m}\) to \(0.064\;\mathrm{m}\). The radial and transverse velocity estimation errors are \(0.032\;\mathrm{m/s}\) and \(0.069\;\mathrm{m/s}\), respectively. Compared with a high-complexity exhaustive \gls{4d} \gls{mle}, the proposed method achieves comparable velocity estimation results while yielding a more accurate localization result when the \gls{4d} \gls{mle} has a practical finite search grid.
\end{abstract}


\section{Introduction}

\IEEEPARstart{N}{ear}-Field (NF) sensing with massive \acrfull{mimo} arrays has recently attracted significant interest in \gls{isac}, since the conventional \gls{ff} plane-wave assumption breaks down when the target lies within the radiative \gls{nf} \cite{zhang_6g_2023, chen_6g_2024, zhao_modeling_2024, xu_distributed_2025}. In \gls{nf}, the received signal exhibits richer spatial information due to spherical wavefront propagation, enabling joint estimation of target location and motion state \cite{wang_near-field_2025-3, lin_isac-enabled_2025-1, miao_near-field_2025-1}.

However, most existing \gls{nf} sensing works focus on localization, while ignoring the potential of estimating velocity and its effect on localization. In \gls{ff}, Doppler is approximated as constant across the array and only represents the radial velocity. In contrast, a moving target in \gls{nf} induces spatially variant Doppler across the array. This spatial Doppler variation enables the estimation of both radial and transverse velocities \cite{wang_near-field_2025-3, lin_isac-enabled_2025-1, miao_near-field_2025-1}. Moreover, neglecting this variation introduces a Doppler mismatch and thus degrades \gls{nf} localization \cite{sakhnini_near-field_2022}.

The potential of \gls{nf} velocity estimation was first highlighted in \cite{wang_near-field_2025-3}, where radial and transverse velocities were estimated through a \gls{2d} \acrfull{mle} assuming known target range and angle. A \acrfull{4d} \gls{mle} was then proposed in \cite{wang_near-field_2025-2}, which can jointly estimate range, angle, radial, and transverse velocities. However, the computational cost of the \gls{4d} nonlinear search with fine grids becomes prohibitive in practice. To reduce the computational complexity, \cite{ebadi_near-field_2025} proposed a two-step subarray-based estimation scheme, which first estimates angle and radial velocity using a subarray under \gls{ff} approximation, and then estimates range and transverse velocity using the full \gls{nf} model. 
This separation lowers complexity, but it also introduces error propagation in the two stages. 
The limitations revealed by the current literature motivate the need for a low-complexity method that exploits spatial Doppler variation in \gls{nf} for velocity estimation while also improving localization.

This paper presents a low-complexity recursive velocity and location estimation framework, which starts from a constant-Doppler coarse localization, then estimates radial and transverse velocities through a closed-form \acrfull{lse}, and performs antenna-dependent Doppler-aware \gls{saf} refinement. We validate the framework through both simulations and measurements, demonstrating its effectiveness against two benchmark methods: a low-complexity constant-Doppler \gls{saf} method as baseline \cite{sakhnini_near-field_2022}, and a high-complexity exhaustive \gls{4d} \gls{mle} as reference \cite{wang_near-field_2025-2}.

The main contributions are summarized as follows:
\begin{itemize}
    \item We formulate the \gls{nf} bistatic Doppler structure for moving targets in \gls{ofdm}-based massive \gls{mimo} sensing and highlight how its spatial variation across the array enables the estimation of joint radial and transverse velocities.
    \item We develop a low-complexity recursive algorithm that alternates between closed-form \gls{lse}-based velocity estimation and Doppler-aware \gls{saf} refinement.
    \item We first validate the proposed framework using simulations. We show that the proposed algorithm significantly improves the estimation accuracy of range, angle, and radial velocity compared to the constant-Doppler baseline, while also enabling transverse velocity estimation. We also show that the proposed method achieves comparable radial/transverse velocity estimates and more accurate localization results than the \gls{4d} \gls{mle} benchmark, which uses a practical finite grid.
    \item We also validate the proposed framework through real-world experiments on a massive \gls{mimo} testbed. Experimental results align well with the simulations. The proposed method achieves velocity estimates comparable to those of the \gls{4d} \gls{mle} and the best localization result among the three methods.
\end{itemize}


\section{System Model}
\label{sec:system_model}

\subsection{Near-Field Geometry and Motion Parameterization}

\begin{figure}[t]
    \centering
    \includegraphics[width=\linewidth]{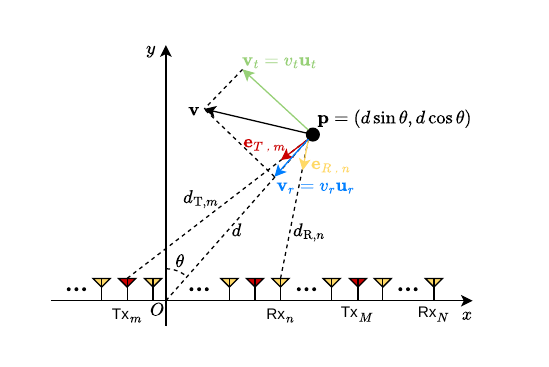}
    \caption{Illustration of the system geometry in \gls{nf}.}
    \label{fig:system_geometry}
\end{figure}

As shown in Fig.~\ref{fig:system_geometry}, we consider an \gls{ofdm}-based massive \gls{mimo} sensing scenario with $M$ \glspl{tx} and $N$ \glspl{rx} located at arbitrary \gls{2d} positions $\mathbf p_{\mathrm T,m}\in\mathbb R^2$, $m \in \left\{1,2,\dots,M\right\}$ and $\mathbf p_{\mathrm R,n}\in\mathbb R^2$, $n \in \left\{1,2,\dots,N\right\}$. A point target is moving in the \gls{nf} sensing region, with position
\(
    \mathbf p
    =
    [\,d\sin\theta,\; d\cos\theta\,]^\top
    \label{eq:target-position}
\)
with respect to the array center,
where \(d>0\) and \(\theta\in(-\pi,\pi]\) denote its range and angle, respectively. For the $(m,n)$-th \gls{tx}-\gls{rx} pair, the bistatic path length and propagation delay are given by
\begin{equation}
    D_{mn}=d_{\mathrm T,m}+d_{\mathrm R,n},
    \quad
    \tau_{mn}=\frac{D_{mn}}{c},
    \label{eq:path-delay}
\end{equation}
where \(d_{\mathrm T,m}=\|\mathbf p_{\mathrm T,m}-\mathbf p\|\) is the distance between the target and the $m$-th \gls{tx}, and \(d_{\mathrm R,n}=\|\mathbf p_{\mathrm R,n}-\mathbf p\|\) is the distance between the target and the $n$-th \gls{rx}, $c$ is the speed of light.

The target velocity is decomposed into orthogonal radial and transverse components as
\begin{equation}
    \mathbf v = v_r \mathbf u_r + v_t \mathbf u_t,
    \label{eq:velocity-decomposition}
\end{equation}
where the radial and transverse unit vectors are defined by 
\(
    \mathbf u_r
    =
    [\,-\sin\theta,\; -\cos\theta\,]^\top
\)
and 
\(
    \mathbf u_t
    =
    [\,-\cos\theta,\; \sin\theta\,]^\top.
    \label{eq:ur-ut}
\)
To characterize the bistatic Doppler of the $(m,n)$-th \gls{tx}-\gls{rx} pair, we first define the radial and transverse bistatic coefficients as
\begin{equation}
    q_{r,mn} = \mathbf u_r^\top (\mathbf e_{\mathrm T,m} + \mathbf e_{\mathrm R,n}),
    \quad
    q_{t,mn} = \mathbf u_t^\top (\mathbf e_{\mathrm T,m} + \mathbf e_{\mathrm R,n}),
    \label{eq:qr-qt-def}
\end{equation}
where \(\mathbf e_{\mathrm T,m} = \frac{\mathbf p_{\mathrm T,m} - \mathbf p}{d_{\mathrm T,m}}\) and \(\mathbf e_{\mathrm R,n} = \frac{\mathbf p_{\mathrm R,n} - \mathbf p}{d_{\mathrm R,n}}\) are the bistatic unit vectors. Then the bistatic Doppler is given by
\begin{equation}
    f_{D,mn}
    =
    \frac{1}{\lambda}
    \left[
        v_r q_{r,mn}
        +
        v_t q_{t,mn}
    \right],
    \label{eq:doppler-qr-qt}
\end{equation}
where $\lambda$ is the wavelength. Equation \eqref{eq:doppler-qr-qt} shows that the bistatic Doppler varies across \gls{tx}-\gls{rx} pairs through the geometry-dependent coefficients $q_{r,mn}$ and $q_{t,mn}$ in the \gls{nf}, which is the key property exploited in this paper. 

\subsection{\gls{ofdm} Signal Model}

We consider one \gls{cpi} consisting of $L$ \gls{ofdm} symbols, indexed by $l\in\{0,1,\dots,L-1\}$. As commonly assumed in conventional radar signal processing, the target position is approximately constant within one \gls{cpi}, i.e., range migration is negligible~\cite{richards2005fundamentals}. The received signal after channel estimation of the $(m,n)$-th \gls{tx}-\gls{rx} pair during one single \gls{ofdm} symbol is then expressed as
\begin{equation}
    h_{mn}[k,l]
    = \beta
    e^{-j2\pi f_k \tau_{mn}}
    e^{j2\pi f_{D,mn} lT_p}
    + w_{mnkl},
    \label{eq:pilot-compensated-observation}
\end{equation}
where \(\beta\in\mathbb{C}\) denotes the target reflection coefficient, $f_k=f_c+k\Delta f$ is the frequency of the $k$-th subcarrier, with $\; k\in\{0,1,\dots,K-1\}$ the subcarrier index, $f_c$ the carrier frequency, and $\Delta f$ the subcarrier spacing. The first phase term in \eqref{eq:pilot-compensated-observation} varies by the bistatic path delay $\tau_{mn}$ derived in \eqref{eq:path-delay}, whereas the second phase term reveals the effect of the bistatic Doppler $f_{D,mn}$ modeled in \eqref{eq:doppler-qr-qt}, together with the \gls{ofdm} symbol duration $T_p$. Moreover, $w_{mnkl}\sim\mathcal{CN}(0,\sigma^2)$ denotes complex additive white Gaussian noise.



\section{Recursive Velocity Estimation and Doppler-Aware Localization Algorithm}

In this section, we propose a low-complexity recursive algorithm for estimating $(d,\theta,v_r,v_t)$ from the received \gls{ofdm} observations in \eqref{eq:pilot-compensated-observation} within one \gls{cpi}. The 
procedure starts from range-Doppler processing for each \gls{tx}-\gls{rx} pair, and a coarse location initialization by constant-Doppler \gls{saf}. The algorithm then alternates between two updates: 1) \gls{lse}-based velocity estimation using the current location estimate, and 2) Doppler-aware localization using the updated velocity estimate. This alternating process is repeated until convergence.

\textit{Range-Doppler Processing:} For each \gls{tx}-\gls{rx} pair, we perform range-Doppler processing through a \gls{2d} Fourier transform on \(h_{mn}[k,l]\) in \eqref{eq:pilot-compensated-observation} as
\begin{equation}
    h^{\mathrm{rd}}_{mn}[\rho,\varphi]
    =
    \frac{1}{KL}
    \sum_{l=0}^{L-1}
    \sum_{k=0}^{K-1}
    h_{mn}[k,l]
    e^{j2\pi k \Delta f \tau_\rho}
    e^{-j2\pi f_{D,\varphi}lT_p},
    \label{eq:rd_processing}
\end{equation}
where $\rho\in\{0,1,\dots,N_\tau-1\}$ denotes the delay-bin index, \(\tau_\rho=\rho/(N_\tau\Delta f)\), and \(\varphi\in\{0,1,\dots,N_D-1\}\) is the Doppler-bin index, \(f_{D,\varphi}=(\varphi-N_D/2)/(N_D T_p)\).

\textit{Coarse Location Initialization:}
We then initialize a coarse location adopting the constant-Doppler \gls{saf} baseline, which assumes that all \gls{tx}-\gls{rx} pairs share one common Doppler bin \cite{sakhnini_near-field_2022}. This Doppler bin is selected from the averaged range-Doppler maps among all the \gls{tx}-\gls{rx} pairs as
\begin{equation}
    \bar\varphi
    =
    \arg\max_{\varphi}
    \frac{1}{MN}
    \sum_{m=1}^{M}\sum_{n=1}^{N}
    \max_{\rho}
    \left|h^{rd}_{mn}[\rho,\varphi]\right|^2.
    \label{eq:common_doppler_bin}
\end{equation}

The corresponding \gls{saf} is formed as the matched filter output:
\begin{equation}
    \mathcal {A}^{\mathrm{CD}}(d,\theta;\bar\varphi)
    \hspace{-0.1cm}=\hspace{-0.1cm}
    \left|
    \sum_{m=1}^{M}\sum_{n=1}^{N}
    h^{rd}_{mn}\!\left[\rho_{mn},\bar\varphi\right]\,
    a_{mn}(d,\theta)
    \right|^2
    \label{eq:constant_doppler_bp}
\end{equation}
where $\rho_{mn}$ denotes the range bin of the strongest range response within the Doppler bin $\bar\varphi$ for the $(m,n)$-th \gls{tx}-\gls{rx} pair, and
\(
    a_{mn}(d,\theta)
    =
    e^{j\frac{2\pi}{\lambda}D_{mn}(d,\theta)}
    \label{eq:test_signal}
\)
is the expected signal at the hypothesized location $(d,\theta)$. The coarse location is then obtained from the peak of the \gls{saf}:
\begin{equation}
    (\hat d^{(0)},\hat\theta^{(0)})
    =
    \arg\max_{d,\theta}\,
    \mathcal {A}^{\mathrm{CD}}(d,\theta;\bar\varphi).
    \label{eq:coarse_location}
\end{equation}

\textit{\gls{lse}-Based Velocity Estimation:}
We apply the coarse location initialized in \eqref{eq:coarse_location} as the input for the first iteration in the recursive process. At the $i$-th iteration, we denote the input location estimate as $(\hat d^{(i-1)},\hat\theta^{(i-1)})$. The corresponding geometry-dependent bistatic coefficients are evaluated as
\begin{equation}
    \hat q_{r,mn}
    =
    q_{r,mn}(\hat d^{(i-1)},\hat\theta^{(i-1)}),
    \quad
    \hat q_{t,mn}
    =
    q_{t,mn}(\hat d^{(i-1)},\hat\theta^{(i-1)}).
    \label{eq:estimated_q}
\end{equation}
We then extract the bistatic Doppler \(\hat f_{D,mn}=(\hat\varphi_{mn}-N_D/2)/(N_D T_p)\) from each \gls{tx}-\gls{rx} pair by selecting the dominant peak of its range-Doppler map:
\begin{equation}
    (\hat\rho_{mn},\hat\varphi_{mn})
    =
    \arg\max_{\rho,\varphi}
    \left|h^{rd}_{mn}[\rho,\varphi]\right|^2.
    \label{eq:doppler_extraction}
\end{equation}
Using \eqref{eq:doppler-qr-qt} and by
stacking the extracted bistatic Doppler from all \gls{tx}-\gls{rx} pairs, we obtain
\begin{equation}
    \hat{\mathbf f}_D
    \approx
    \frac{1}{\lambda}\,
    \mathbf \Psi(\hat d^{(i-1)},\hat\theta^{(i-1)}) [v_r,
        v_t]\mathsf{T}
    \label{eq:stacked_doppler_model}
\end{equation}
with \(\mathbf \Psi\) stacking the bistatic coefficients from \eqref{eq:estimated_q}.
The radial and transverse velocities are then estimated using a closed-form \gls{lse}:
\begin{equation}
[\hat v_r^{(i)},
        \hat v_t^{(i)}] ^\mathsf{T}
    =
    \lambda
    \left(
        \mathbf \Psi^\mathsf{T}\mathbf \Psi
    \right)^{-1}
    \mathbf \Psi^\mathsf{T}
    \hat{\mathbf f}_D.
    \label{eq:lse_velocity}
\end{equation}

\textit{Doppler-Aware Localization:}
Given the velocity estimate $(\hat v_r^{(i)},\hat v_t^{(i)})$ at the $i$-th iteration, we compute the corresponding bistatic Doppler \(\hat f_{D,mn}^{(i)}\) using \eqref{eq:doppler-qr-qt} for each \gls{tx}-\gls{rx} pair.
The Doppler bin is then selected as the closest bin to the predicted bistatic Doppler:
\begin{equation}
    \varphi_{mn}^{(i)}
    =
    \arg\min_{\varphi}
    \left|
        f_{D,\varphi}
        -
        \hat f_{D,mn}^{(i)}
    \right|.
    \label{eq:predicted_doppler_bin}
\end{equation}
Then, only the strongest range response within the selected Doppler bin is retained:
\begin{equation}
    \rho_{mn}^{(i)}
    =
    \arg\max_{\rho}
    \left|h^{rd}_{mn}[\rho,\varphi_{mn}^{(i)}]\right|^2.
    \label{eq:selected_range_bin}
\end{equation}
Using these antenna-dependent range-Doppler selections, the Doppler-aware \gls{saf} is formed as
\begin{equation}
    \mathcal {A}^{(i)}(d,\theta)
    =
    \left|
    \sum_{m=1}^{M}\sum_{n=1}^{N}
    h^{rd}_{mn}\!\left[\rho_{mn}^{(i)},\varphi_{mn}^{(i)}\right]\,
    a_{mn}(d,\theta)
    \right|^2.
    \label{eq:doppler_aware_bp}
\end{equation}
The location estimate is then refined as
\begin{equation}
    (\hat d^{(i)},\hat\theta^{(i)})
    =
    \arg\max_{d,\theta}\,
    \mathcal {A}^{(i)}(d,\theta).
    \label{eq:location_update}
\end{equation}

\textit{Recursive Refinement and Termination Criterion:}
The framework recursively applies the velocity update in \eqref{eq:lse_velocity} and the location refinement in \eqref{eq:location_update}. The recursion is terminated when the updates of all four unknown parameters are below their predefined thresholds, i.e.,
\begin{equation}
\begin{aligned}
    &\left|\hat d^{(i+1)}-\hat d^{(i)}\right| < \epsilon_d, \quad \;
    \left|\hat\theta^{(i+1)}-\hat\theta^{(i)}\right| < \epsilon_\theta,\\
    &\left|\hat v_r^{(i+1)}-\hat v_r^{(i)}\right| < \epsilon_{v_r}, \quad
    \left|\hat v_t^{(i+1)}-\hat v_t^{(i)}\right| < \epsilon_{v_t}.
\end{aligned}
\label{eq:recursive_convergence}
\end{equation}

\section{Simulation Results}

\begin{table}[t]
\caption{Massive \gls{mimo} system parameters.}
\begin{center}
\begin{tabular}{|l|l|l|}
\hline
\textbf{System parameter} & \textbf{Symbol} & \textbf{Value}   \\ \hline
Center frequency           & $f_c$           & \SI{3.5}{\GHz}   \\ \hline
Subcarrier spacing         & $\Delta f$      & \SI{180}{\kHz}   \\ \hline
Bandwidth                  & BW              & \SI{18}{\MHz}    \\ \hline
Sampling rate              & $f_s$           & \SI{2}{\kHz}     \\ \hline
Symbol duration            & $T_p$             & \SI{0.5}{\ms}    \\ \hline
Number of subcarriers      & K               & 100              \\ \hline
Number of symbols         & L                & 1024
\\ \hline
Number of transmitter antennas & M           & 4              \\ \hline
Number of receiver antennas& N               & 64                \\ \hline
Antenna spacing            & $\Delta a$      & \SI{0.07}{\m}     \\ \hline
\end{tabular}
\label{tab:mamimo_param}
\end{center}
\vspace{-0.2cm}
\end{table}

In this section, we evaluate the proposed algorithm through numerical simulations and compare it with two benchmarks: the constant-Doppler \gls{saf} \cite{sakhnini_near-field_2022} and the exhaustive \gls{4d} grid-search \gls{mle} \cite{wang_near-field_2025-2}. For consistency, the simulation parameters match the experimental settings listed in Table~\ref{tab:mamimo_param}. The \glspl{tx} and \glspl{rx} are arranged as a \gls{ula} to clearly illustrate the spatial Doppler variation. The target state is fixed to \(d=4 \; m\), \(\theta=15\,\degree\), \(v_r=1 \; m/s\), and \(v_t=1 \; m/s\). The simulations use an \gls{snr} of \(20\;\mathrm{dB}\).

\begin{table*}[t]
\caption{Simulation comparison of estimation errors among the three methods.}
\begin{center}
\begin{tabular}{|l|c|c|c|c|c|}
\hline
Method & $e_d=|\hat d-d|$ [m] & $e_\theta=|\hat \theta-\theta|$ [$\degree$] & $e_{v_r}=|\hat v_r-v_r|$ [m/s] & $e_{v_t}=|\hat v_t-v_t|$ [m/s] & $e_{\mathrm{loc}} = \|\hat{\mathbf p}-\mathbf p\|_2$ [m] \\ \hline
Constant-Doppler \gls{saf} & 0.271825 & 3.669909 & 0.952756 & N/A & 0.367491  \\ \hline
Proposed recursive method & 0.007734 & 0.002521 & 0.014495 & 0.000236 & 0.007736 \\ \hline
Exhaustive 4D \gls{mle} & 0.043590 & 0.146906 & 0.002564 & 0.012821 & 0.044793 \\ \hline
\end{tabular}
\label{tab:simulation_comparison}
\end{center}
\vspace{-0.6cm}
\end{table*}


\begin{figure}[t]
    \centering
    \includegraphics[width=1\linewidth]{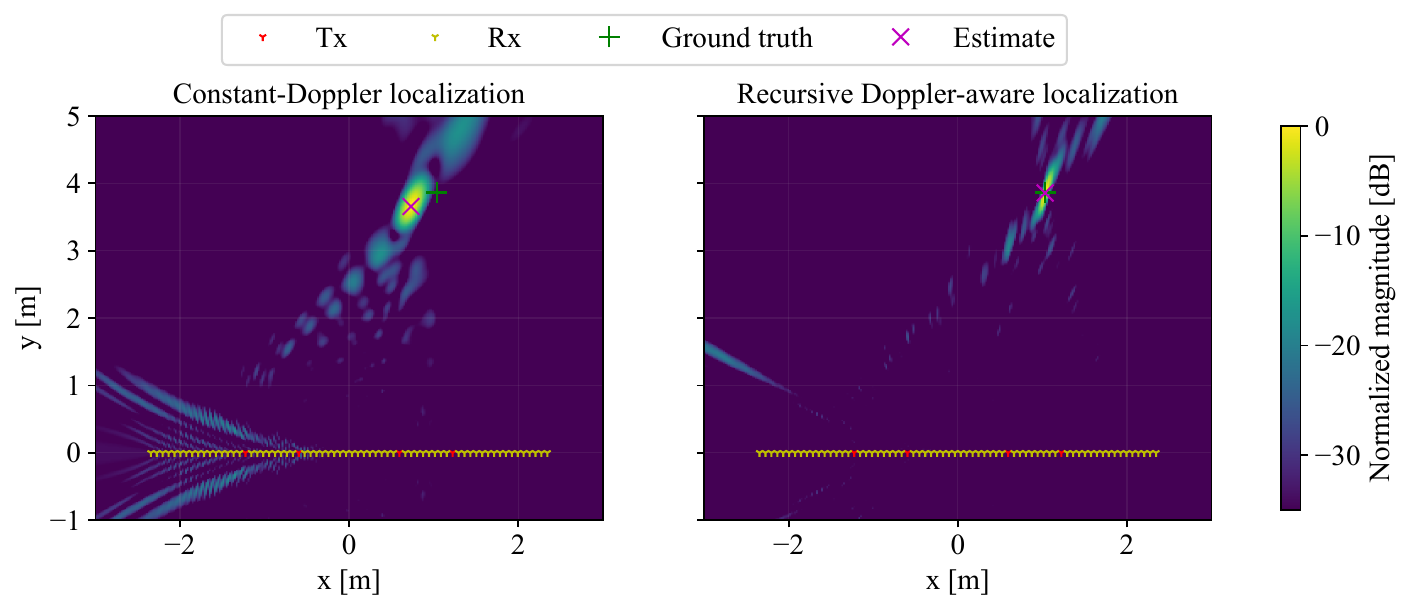}
    \caption{\glspl{saf} of the constant-Doppler baseline and the proposed recursive Doppler-aware localization after two iterations using simulation data.}
    \label{fig:Localization_maps_constant_vs_recursive}
    \vspace{-0.2cm}
\end{figure}

Table~\ref{tab:simulation_comparison} presents the estimation errors of the three methods. The constant-Doppler \gls{saf} yields a coarse location estimate, with range, angle, and overall localization errors of \(0.272\;\mathrm{m}\), \(3.67\,\degree\), and \(0.367\;\mathrm{m}\), respectively. Since this benchmark assumes one common Doppler for all \gls{tx}-\gls{rx} pairs, it only estimates the radial velocity and cannot recover the transverse velocity. The resulting radial velocity error is \(0.953\;\mathrm{m/s}\). In contrast, the proposed recursive method significantly improves both localization and velocity estimation. After only two iterations, the range and angle errors are reduced to \(7.734\times10^{-3}\;\mathrm{m}\) and \(2.521\times10^{-3}\;\degree\), resulting in a localization error of \(7.736\times10^{-3}\;\mathrm{m}\). Meanwhile, the radial and transverse velocity errors decrease to \(0.014\;\mathrm{m/s}\) and \(2.36\times10^{-4}\;\mathrm{m/s}\), respectively. As a high-complexity benchmark, we assume that a coarse search has already been performed before applying the exhaustive grid search in the \gls{4d} \gls{mle} to reduce the simulation burden. So the \gls{4d} \gls{mle} is implemented over a \(40\times40\times40\times40\) grid, with a search range of $0.2$ centered around the ground truth for each unknown parameter.
The \gls{4d} \gls{mle} also achieves accurate joint estimation for $(d,\theta,v_r,v_t)$, with errors of \(0.0436\;\mathrm{m}\), \(0.147\,\degree\), \(2.564\times10^{-3}\;\mathrm{m/s}\), and \(0.0128\;\mathrm{m/s}\). However, its localization accuracy is lower than that of the proposed recursive method, since its grid resolutions cannot be infinitesimally small in practice.


Fig.~\ref{fig:Localization_maps_constant_vs_recursive} further shows the \glspl{saf} resulting from the constant-Doppler baseline and the proposed method. The constant-Doppler \gls{saf} produces a broadened and biased peak, indicating a loss of coherent focus due to the Doppler mismatch, which also explains the relatively large localization error reported in Table~\ref{tab:simulation_comparison}. After two iterations, the proposed Doppler-aware method yields a much sharper and better-focused peak around the true target location. 

We briefly discuss the computational complexity of the three methods. The constant-Doppler \gls{saf} has the lowest complexity, since it requires only one \gls{2d} search over $(d,\theta)$ after range-Doppler processing. Compared with the constant-Doppler \gls{saf}, the additional cost of the proposed recursive method mainly comes from extracting bistatic Doppler per \gls{tx}-\gls{rx} pair, and performing a few iterations on solving the $2\times2$ closed-form \gls{lse} and the Doppler-aware \gls{saf} refinements. As for the \gls{4d} \gls{mle}, its complexity is substantially higher because the search dimension increases from two to four.

Overall, the simulation results confirm that the proposed recursive framework provides a favorable tradeoff between estimation accuracy and computational complexity. It outperforms the low-complexity constant-Doppler baseline and exceeds the performance of the \gls{4d} \gls{mle} implemented with finite grid resolutions.



\section{Experimental Validation}

We validate the proposed method using measurement data collected on the KU Leuven massive \gls{mimo} testbed. The comparison again includes the proposed recursive algorithm with the constant-Doppler \gls{saf} and the exhaustive \gls{4d} \gls{mle}.

\begin{figure}[t]
    \centering
    \includegraphics[width=0.9\linewidth]{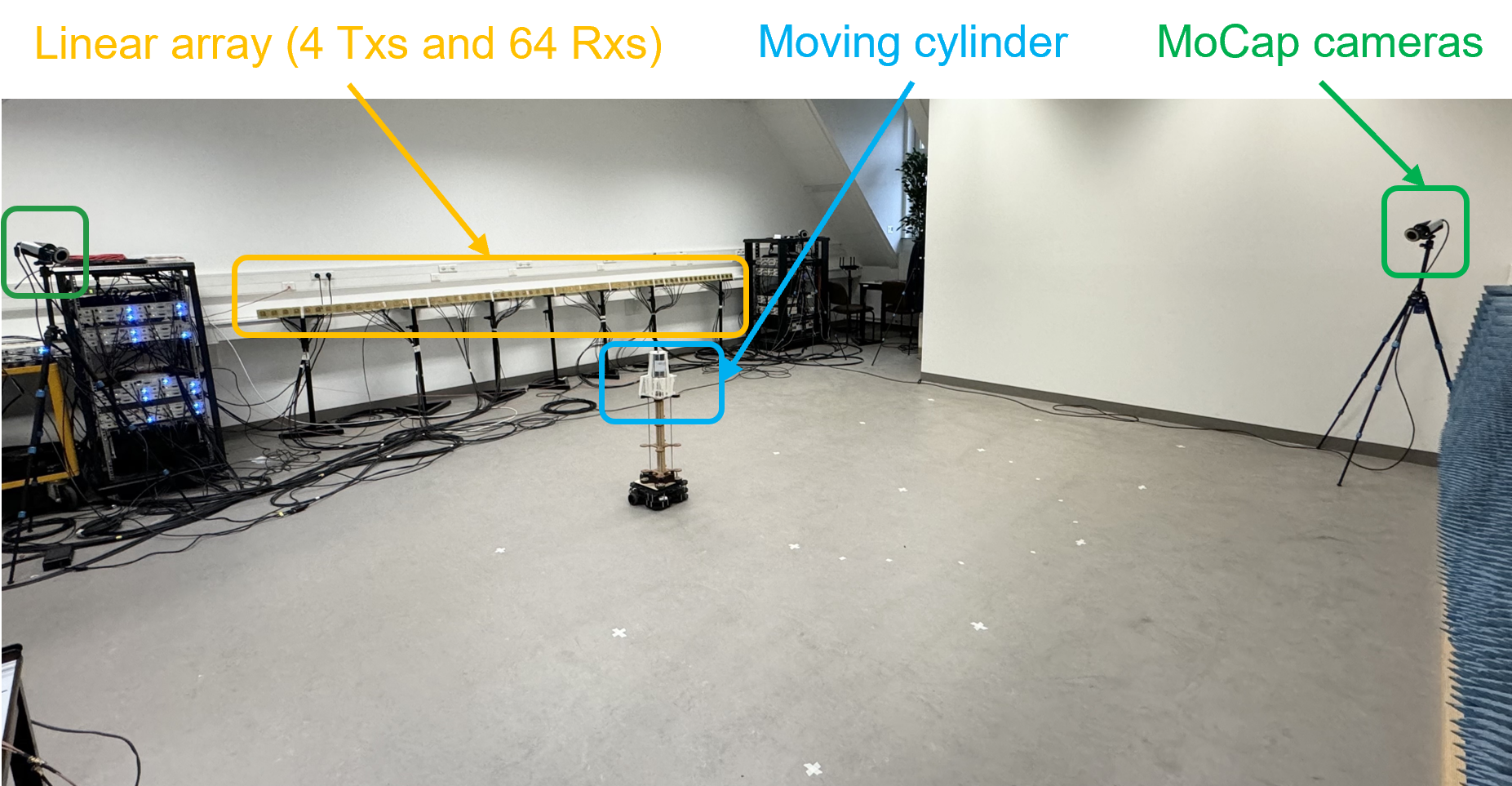}
    \caption{Experimental setup for real-world validation.}
    \label{fig:measurement_setup}
    \vspace{-0.2cm}
\end{figure}

The experimental setup is shown in Fig.~\ref{fig:measurement_setup}. It consists of a linear array with $4$ \glspl{tx} and $64$ \glspl{rx}, a moving cylindrical target in front of the array, and a \gls{mocap} camera system that provides ground-truth target locations. The massive \gls{mimo} system parameters are listed in Table~\ref{tab:mamimo_param}.
For one \gls{cpi}, the measured data are reshaped into a data cube of size $M\times N\times K\times L$. The \gls{mocap} measurements over the entire trajectory are downsampled to the \gls{cpi} timestamps, and the corresponding ground-truth target velocity is obtained from the \gls{mocap} trajectory by finite differences.
Before applying the velocity estimation and localization algorithm, measured data are calibrated to compensate for frequency-dependent and antenna-dependent phase offsets due to hardware imperfections \cite{zhang_robust_2026}. This step is essential for restoring phase coherence across the array and enabling accurate \gls{nf} localization.

\begin{table*}[t]
\caption{Experimental comparison of estimation errors among the three methods.}
\begin{center}
\begin{tabular}{|l|c|c|c|c|c|}
\hline
Method & $e_d=|\hat d-d|$ [m] & $e_\theta=|\hat \theta-\theta|$ [$\degree$] & $e_{v_r}=|\hat v_r-v_r|$ [m/s] & $e_{v_t}=|\hat v_t-v_t|$ [m/s] & $e_{\mathrm{loc}} = \|\hat{\mathbf p}-\mathbf p\|_2$ [m] \\ \hline
Constant-Doppler \gls{saf} & 0.267165 & 0.556227 & 0.032646 & N/A & 0.267584  \\ \hline
Proposed recursive method & 0.063596 & 0.215432 & 0.032378 & 0.068892 & 0.063898 \\ \hline
Exhaustive 4D \gls{mle} & 0.100000 & 3.378961 & 0.100000 & 0.043590 & 0.142864 \\ \hline
\end{tabular}
\label{tab:experimental_comparison}
\end{center}
\vspace{-0.6cm}
\end{table*}

\begin{figure}[t]
    \centering
    \includegraphics[width=1\linewidth]{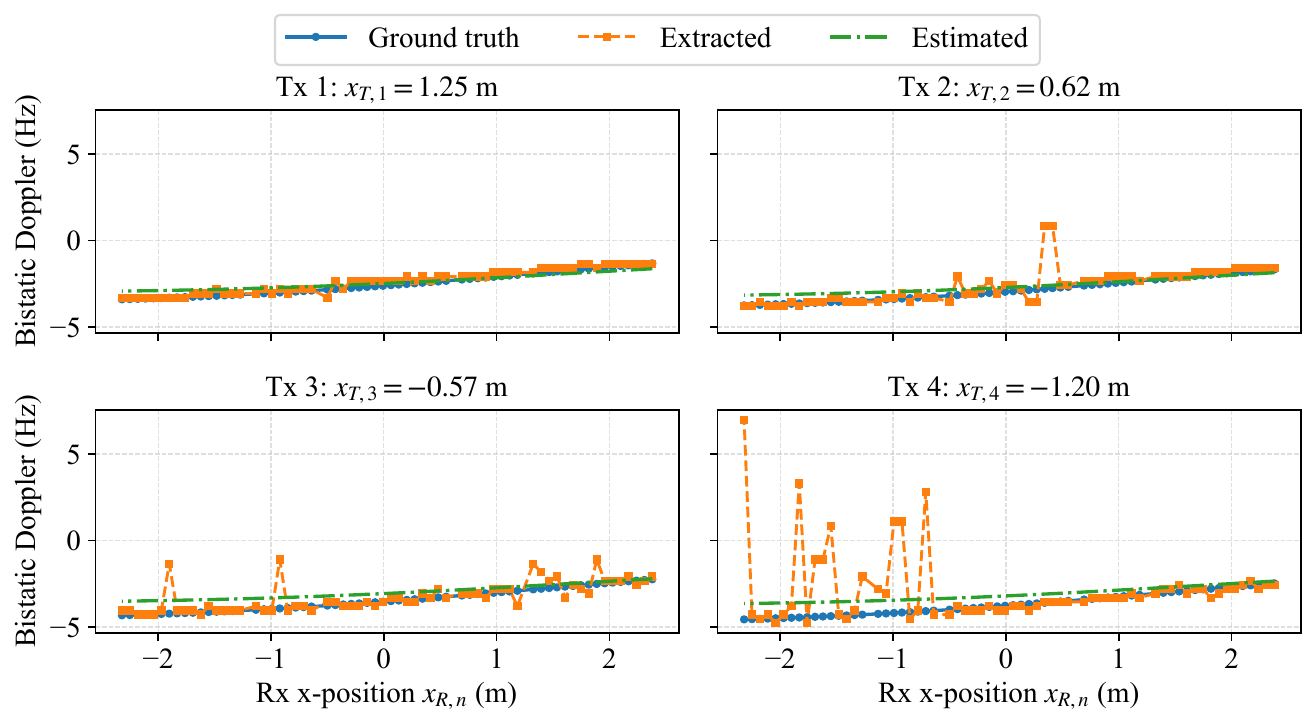}
    \caption{Measured and estimated bistatic Doppler across the Rx aperture for the four \glspl{tx}, using the proposed recursive method.}
    \label{fig:Bistatic_Doppler_Comparison_frame_005}
    \vspace{-0.2cm}
\end{figure}

\begin{figure}[t]
    \centering
    \includegraphics[width=1\linewidth]{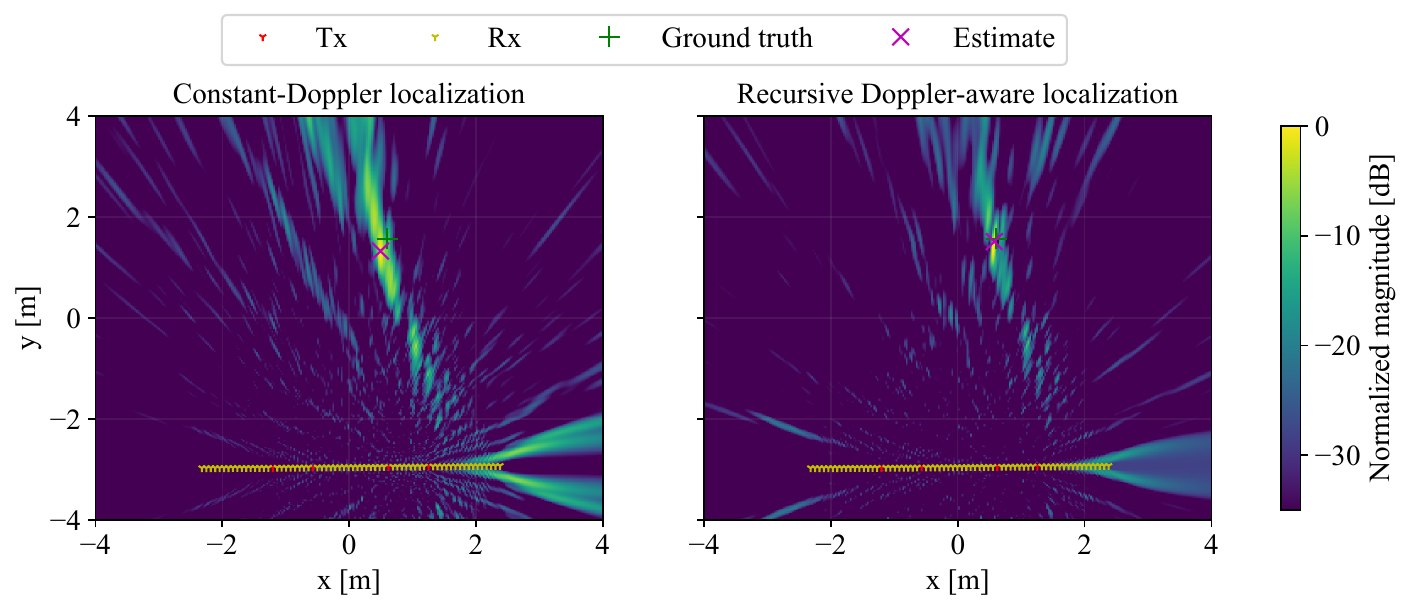}
    \caption{\glspl{saf} of the constant-Doppler baseline and the proposed recursive Doppler-aware localization after two iterations using measurement data.}
    \label{fig:Localization_maps_constant_vs_recursive_frame_005}
    \vspace{-0.2cm}
\end{figure}

Table~\ref{tab:experimental_comparison} indicates that the experimental results align well with the simulation observations. The constant-Doppler \gls{saf} is sufficient only for coarse initialization, whereas incorporating the antenna-dependent bistatic Doppler significantly improves localization performance. In particular, the proposed recursive method reduces the localization error from \(0.268\;\mathrm{m}\) to \(0.0639\;\mathrm{m}\), and it also enables the transverse velocity estimation. The exhaustive \gls{4d} \gls{mle} also does not achieve the same localization accuracy as the proposed recursive method under the practical finite-grid setting.

Fig.~\ref{fig:Bistatic_Doppler_Comparison_frame_005} shows the extracted bistatic Doppler across the \gls{rx} aperture for all four \glspl{tx}. The measured Doppler generally follows the spatial variation of the ground truth, confirming the presence of \gls{nf} Doppler variation in the measurements. However, several visible spikes can be observed, indicating noisy bistatic Doppler extraction in practice that may degrade the velocity estimates. Nevertheless, the bistatic Doppler calculated from the final velocity estimate of the proposed recursive algorithm still captures the overall trend of the ground-truth curves, showing that the dominant spatial Doppler structure is successfully recovered despite imperfect observations.

The \glspl{saf} are shown in Fig.~\ref{fig:Localization_maps_constant_vs_recursive_frame_005}. Under the constant-Doppler assumption, the \gls{saf} is visibly defocused, and its peak is displaced from the ground-truth target position. After two iterations, the peak becomes sharper and shifts closer to the true location, consistent with the error reduction reported in Table~\ref{tab:experimental_comparison} and the insights in the previous simulation. Moreover, the recursive framework provides additional robustness against the noisy Doppler extraction in Fig.~\ref{fig:Bistatic_Doppler_Comparison_frame_005}. By using the velocity estimate to guide the antenna-dependent Doppler selection, it mitigates the impact of Doppler outliers and stabilizes the localization process on measured data.

Overall, the experimental results are consistent with the simulation observations. 
The proposed recursive method jointly estimates both radial and transverse velocities, and achieves the best localization result among the three methods, demonstrating an effective balance between complexity, robustness, and estimation performance in practical \gls{nf} scenarios.

\section{Conclusion}
\label{sec: conclusion}

This paper studied \gls{nf} velocity estimation and Doppler-aware localization for \gls{ofdm}-based massive \gls{mimo} sensing. We showed the \gls{nf} spatial Doppler variation across \gls{tx}-\gls{rx} pairs parameterized by the target location and its radial and transverse velocities. Based on this structure, we developed a low-complexity recursive algorithm that combines closed-form \gls{lse}-based velocity estimation with Doppler-aware \gls{saf} localization refinement. Compared with the constant-Doppler baseline, the proposed method significantly improves localization while additionally enabling transverse velocity estimation. In simulation, it reduces the localization error from \(0.367\;\mathrm{m}\) to \(7.74\times10^{-3}\;\mathrm{m}\), and in experiments from \(0.268\;\mathrm{m}\) to \(0.0639\;\mathrm{m}\). The results also show that the proposed method can outperform the high-complexity exhaustive \gls{4d} \gls{mle} in localization, when the \gls{4d} \gls{mle} has practical finite grid resolutions. Overall, the proposed framework provides an effective and robust tradeoff between complexity and estimation performance for practical \gls{nf} sensing.

\balance
\bibliographystyle{IEEEtran}
\bibliography{references,ref_manual}

@book{richards2005fundamentals,
  title={Fundamentals of radar signal processing},
  author={Richards, Mark A and others},
  volume={1},
  year={2005},
  publisher={Mcgraw-hill New York}
}

@inproceedings{zhang_robust_2026,
	title = {Robust {Localization} in {OFDM}-{Based} {Massive} {MIMO} through {Phase} {Offset} {Calibration}},
	url = {https://ieeexplore.ieee.org/abstract/document/11366003},
	doi = {10.1109/JCS69321.2026.11366003},
	urldate = {2026-04-29},
	booktitle = {2026 {IEEE} 6th {International} {Symposium} on {Joint} {Communications} \& {Sensing} ({JC}\&{S})},
	author = {Zhang, Qing and Sakhnini, Adham and Beerten, Robbert and Xiong, Haoqiu and Cui, Zhuangzhuang and Miao, Yang and Pollin, Sofie},
	month = jan,
	year = {2026},
	pages = {1--6},
}

@article{lin_isac-enabled_2025-1,
	title = {{ISAC}-{Enabled} {Near}-{Field} {High}-{Mobility} {UAV} {Localization}: {Analysis}, {Bounds} and {Algorithm}},
	issn = {1557-9603},
	shorttitle = {{ISAC}-{Enabled} {Near}-{Field} {High}-{Mobility} {UAV} {Localization}},
	url = {https://ieeexplore.ieee.org/document/11139105},
	doi = {10.1109/TAES.2025.3602379},
	urldate = {2025-11-16},
	journal = {IEEE Transactions on Aerospace and Electronic Systems},
	author = {Lin, Luning and Wang, Mingxing and Shi, Zhiguo and Greco, Maria Sabrina and Gini, Fulvio},
	year = {2025},
	pages = {1--15},
}

@article{miao_near-field_2025-1,
	title = {Near-{Field} {High}-{Speed} {User} {Sensing} in {Wideband} {mmWave} {Communications}: {Algorithms} and {Bounds}},
	volume = {73},
	issn = {1941-0476},
	shorttitle = {Near-{Field} {High}-{Speed} {User} {Sensing} in {Wideband} {mmWave} {Communications}},
	url = {https://ieeexplore.ieee.org/document/10857463},
	doi = {10.1109/TSP.2025.3535691},
	urldate = {2025-11-16},
	journal = {IEEE Transactions on Signal Processing},
	author = {Miao, Hongxia and Peng, Mugen},
	year = {2025},
	pages = {919--935},
}

@article{zhao_modeling_2024,
	title = {Modeling and {Analysis} of {Near}-{Field} {ISAC}},
	volume = {18},
	issn = {1941-0484},
	url = {https://ieeexplore.ieee.org/document/10498098},
	doi = {10.1109/JSTSP.2024.3386054},
	number = {4},
	urldate = {2025-11-16},
	journal = {IEEE Journal of Selected Topics in Signal Processing},
	author = {Zhao, Boqun and Ouyang, Chongjun and Liu, Yuanwei and Zhang, Xingqi and Poor, H. Vincent},
	month = may,
	year = {2024},
	pages = {678--693},
}

@article{xu_distributed_2025,
	title = {Distributed {Signal} {Processing} for {Extremely} {Large}-{Scale} {Antenna} {Array} {Systems}: {State}-of-the-{Art} and {Future} {Directions}},
	volume = {19},
	issn = {1941-0484},
	shorttitle = {Distributed {Signal} {Processing} for {Extremely} {Large}-{Scale} {Antenna} {Array} {Systems}},
	url = {https://ieeexplore.ieee.org/document/10883023},
	doi = {10.1109/JSTSP.2025.3541386},
	number = {2},
	urldate = {2025-11-10},
	journal = {IEEE Journal of Selected Topics in Signal Processing},
	author = {Xu, Yanqing and Larsson, Erik G. and Jorswieck, Eduard A. and Li, Xiao and Jin, Shi and Chang, Tsung-Hui},
	month = mar,
	year = {2025},
	pages = {304--330},
}

@inproceedings{ebadi_near-field_2025,
	address = {Surrey, United Kingdom},
	title = {Near-{Field} {Source} {Localization} and {Velocity} {Estimation} using an {Extremely} {Large} {Antenna} {Array}},
	copyright = {https://doi.org/10.15223/policy-029},
	isbn = {978-1-6654-7776-5},
	url = {https://ieeexplore.ieee.org/document/11143351/},
	doi = {10.1109/SPAWC66079.2025.11143351},
	language = {en},
	urldate = {2025-11-03},
	booktitle = {2025 {IEEE} 26th {International} {Workshop} on {Signal} {Processing} and {Artificial} {Intelligence} for {Wireless} {Communications} ({SPAWC})},
	publisher = {IEEE},
	author = {Ebadi, Zohreh and Molaei, Amir Masoud and Babar Abbasi, Muhammad Ali and Cotton, Simon and Tukmanov, Anvar and Yurduseven, Okan},
	month = jul,
	year = {2025},
	pages = {1--5},
}

@article{wang_near-field_2025-2,
	title = {Near-{Field} {Localization} and {Sensing} {With} {Large}-{Aperture} {Arrays}: {From} signal modeling to processing},
	volume = {42},
	issn = {1558-0792},
	shorttitle = {Near-{Field} {Localization} and {Sensing} {With} {Large}-{Aperture} {Arrays}},
	url = {https://ieeexplore.ieee.org/abstract/document/10934790},
	doi = {10.1109/MSP.2024.3486471},
	number = {1},
	urldate = {2025-10-31},
	journal = {IEEE Signal Processing Magazine},
	author = {Wang, Zhaolin and Ramezani, Parisa and Liu, Yuanwei and Björnson, Emil},
	month = jan,
	year = {2025},
	pages = {74--87},
}

@article{wang_near-field_2025-3,
	title = {Near-{Field} {Velocity} {Sensing} and {Predictive} {Beamforming}},
	volume = {74},
	issn = {0018-9545, 1939-9359},
	url = {http://arxiv.org/abs/2311.09888},
	doi = {10.1109/TVT.2024.3454481},
	number = {1},
	urldate = {2025-04-05},
	journal = {IEEE Transactions on Vehicular Technology},
	author = {Wang, Zhaolin and Mu, Xidong and Liu, Yuanwei},
	month = jan,
	year = {2025},
	note = {arXiv:2311.09888 [cs]},
	pages = {1806--1810},
}

@article{zhang_6g_2023,
	title = {{6G} {Wireless} {Communications}: {From} {Far}-{Field} {Beam} {Steering} to {Near}-{Field} {Beam} {Focusing}},
	volume = {61},
	issn = {1558-1896},
	shorttitle = {{6G} {Wireless} {Communications}},
	url = {https://ieeexplore.ieee.org/document/10068140/authors#authors},
	doi = {10.1109/MCOM.001.2200259},
	number = {4},
	urldate = {2024-11-13},
	journal = {IEEE Communications Magazine},
	author = {Zhang, Haiyang and Shlezinger, Nir and Guidi, Francesco and Dardari, Davide and Eldar, Yonina C.},
	month = apr,
	year = {2023},
	note = {Conference Name: IEEE Communications Magazine},
	pages = {72--77},
}

@article{chen_6g_2024,
	title = {{6G} {Localization} and {Sensing} in the {Near} {Field}: {Features}, {Opportunities}, and {Challenges}},
	volume = {31},
	issn = {1558-0687},
	shorttitle = {{6G} {Localization} and {Sensing} in the {Near} {Field}},
	url = {https://ieeexplore.ieee.org/abstract/document/10529957},
	doi = {10.1109/MWC.011.2300359},
	number = {4},
	urldate = {2024-11-05},
	journal = {IEEE Wireless Communications},
	author = {Chen, Hui and Keskin, Musa Furkan and Sakhnini, Adham and Decarli, Nicolò and Pollin, Sofie and Dardari, Davide and Wymeersch, Henk},
	month = aug,
	year = {2024},
	note = {Conference Name: IEEE Wireless Communications},
	pages = {260--267},
}

@article{sakhnini_near-field_2022,
	title = {Near-{Field} {Coherent} {Radar} {Sensing} {Using} a {Massive} {MIMO} {Communication} {Testbed}},
	volume = {21},
	issn = {1558-2248},
	url = {https://ieeexplore.ieee.org/abstract/document/9707730},
	doi = {10.1109/TWC.2022.3148035},
	number = {8},
	urldate = {2024-07-25},
	journal = {IEEE Transactions on Wireless Communications},
	author = {Sakhnini, Adham and De Bast, Sibren and Guenach, Mamoun and Bourdoux, André and Sahli, Hichem and Pollin, Sofie},
	month = aug,
	year = {2022},
	note = {Conference Name: IEEE Transactions on Wireless Communications},
	pages = {6256--6270},
}

\end{document}